\documentclass[preprint,aps]{revtex4}
\usepackage{graphicx}
\usepackage{dcolumn}
\usepackage{bm}

\begin{document}
\preprint{APS/123-QED}

\title{Wave Mechanics of a Two Wire Atomic Beamsplitter}
\author{Daniele C. E. Bortolotti}
\author{John L. Bohn}
 \email{bohn@murphy.colorado.edu}
\affiliation{JILA and Department of Physics,
University of Colorado, Boulder, CO 80309-0440}

\date{\today}

\begin{abstract}

We consider the problem of an atomic beam propagating quantum
mechanically through an atom
beam splitter.  Casting the problem in an adiabatic representation
(in the spirit of the Born-Oppenheimer approximation in molecular
physics) sheds light on explicit effects due to non-adiabatic passage
of the atoms through the splitter region.
 We are thus able to probe the fully three dimensional structure
of the beam splitter, gathering quantitative information about mode-mixing,
splitting ratios,and reflection and transmission probabilities.
 
\end{abstract}

\maketitle
\narrowtext

\section{Introduction}

Continuing advances in the production and manipulation of atomic
Bose-Einstein condensates (BEC's) are tending toward novel applications in
interferometry.  BEC's can now be produced {\it in situ} on surfaces
\cite{Hansel,Kasper,Ott}, 
making them ready for loading into ``interferometer-on-a-chip''
micro-structures.  Being in close proximity to the chip, the
atoms are subject to control via magnetic fields generated by wires on
the chip. Because of their coherence and greater brightness, Bose-
condensed atoms are expected to improve upon previous accomplishments
with thermal atoms, such as
neutral atom guiding, \cite{Schmiedmayer,muller,dekker,Sauer,Engels}, 
switching \cite{muller2}, and multi-mode beamsplitting 
\cite{Cassettari,muller3}.  Studies of propagation of BEC's through waveguide
structures are also underway \cite{leanhardt}.

While the BEC is created in its lowest transverse mode in the guiding 
potential, keeping it in this mode as it travels through the chip remains
remains a significant technical challenge.  For example, it appears
that inhomogeneities in the guiding wires produce field fluctuations
that can break up the condensate wave function \cite{leanhardt,schwindt}.  
Additionally, the very act of splitting a condensate into two paths implies 
a transverse pull on the condensate that can excite higher modes.  Ideally, 
the condensate propagates sufficiently slowly that, once in its
lowest mode, it follows adiabatically into the lowest mode of the
split condensate.   The criterion for this to happen, roughly, is that
the condensate 
velocity in the direction of motion be less than $L \omega$, where $L$
is a characteristic length scale over which the beam is split, and
$\omega$ is a characteristic frequency of transverse oscillation in the
guiding potential.  Reference \cite{zozulya} has verified this conclusion
numerically, in a two-dimensional model that varies the transverse
potential in time, at a rate equivalent to the passage of the moving
condensate through a beam splitter.  Populating higher modes can reduce
fringe contrast, thus spoiling the operation of an
interferometer. Diffraction has also been pointed out to have negative
effects on guiding in general \cite{Jaask}.

Moving too slowly through the beam splitter is, however, potentially
dangerous because of threshold scattering behavior in a varying potential.
In one dimension, a wave incident on a scattering potential is reflected
with unit probability in the limit of zero collision energy 
\cite{sadeghpour}.  This same kind of ``quantum reflection'' will
be generically present in beam splitters as well, where scattering
can occur from changes in the transverse potential as the
longitudinal coordinate varies.  Reflection upon entering the beam splitter
region can prove devastating for potential applications such as a
Sagnac interferometer.

Both aspects of instability in an atom interferometer can be expressed
in terms of quantum mechanical scattering theory of the atoms from
the guiding potential.  Specifically, a condensate entering a
beam splitter in arm $a$ and in transverse mode $m$ possesses a
scattering amplitude $S_{am,a'm'}$ for exiting in arm $a'$ in mode $m'$.
In this paper we therefore cast the general problem of beam splitting
in terms of scattering theory.  For the time being we restrict our
attention to  the {\it linear} scattering problem, and therefore
implicitly consider the regime of weak inter-atomic interactions.
This is suitable, since the basic question we raise is the effect of
wave mechanical propagation on the atoms.  Note that the weakly
interacting atom limit is achieved with small atom number,
in which case number fluctuations may be problematic \cite{zozulya2}.
Alternatively, this limit is reached at low atom density, which is achieved
for a BEC that has expanded longitudinally for some time before entering
the beam splitter region.

Restricting our attention to the linear Schr\"{o}dinger equation
opens up a host of   
powerful theoretical tools that have been developed in the context of
atomic scattering.  In the present instance, given the dominant role
of non-adiabatic effects, the tool of most use is the adiabatic
representation.  This is analogous to the Born-Oppenheimer approximation
in molecular physics \cite{Levine}.  Specifically, we freeze the value 
of the longitudinal
coordinate $z$ and solve the remaining 2-dimensional Schr\"{o}dinger equation
in $x$-$y$. The resulting $z$-dependent energy spectrum represents a set
of potential curves for following the remaining motion in $z$.  
This general approach has been applied previously to a model situation 
in which the transverse potential is gently contracted or expanded
\cite{Jaask,Jaask2}; here we extend it to realistic waveguide geometries.

This representation has obvious appeal for the problem at hand, since
in this level of approximation it is assumed that the atoms move 
infinitely slowly through the beam splitter.  It is, however, an
exact representation of scattering theory, and the leftover non-adiabatic
corrections, arising from finite propagation velocity, can be explicitly 
incorporated. We will see that non-adiabatic effects have a strong
influence on beamsplitters based on experimentally realistic
parameters.  The effects of excitation of higher transverse modes and
of reflection from the beam splitter therefore have a fairly simple
interpretation  
in these explicit nonadiabticites.  In addition, the successive solution of a 
set of two-dimensional problems in transverse coordinates $x-y$,
followed by a coupled-channel  
calculation in $z$, is less numerically intensive than
than determining the full 3-dimensional solution all at once. Indeed,
this is why adiabatic representations have found widespread use in
chemical physics.  Larger
problems, more closely resembling experimental beam splitters,
can therefore be handled.
This paper is organized as follows: In section II we introduce the
model, describing how the beamsplitter works in general terms
and outlining the theoretical methods used in the paper,
introducing the main ideas about the adiabatic formalism.
In section III we present the
results obtained from our study, with a focus on the description of
the theory itself, and how its different components relate to the
physics of the problem.

\section{Model}

The salient characteristics of a two wire atomic beam-splitter can be
realized in the following way: a guiding potential is generated by the
magnetic field 
due to two parallel current carrying wires and an additional
bias field perpendicular to them. By suitably decreasing the
bias field or the distance
between the wires, it is possible to decrease the separation between
two minima, and thus increase the  probability for the atoms to
tunnel from one to another.

\subsection{The Beam Splitting Potential.}

We start by considering the magnetic field generated by two infinitely
long parallel wires lying on a substrate, each carrying a current
$I$ in the $\hat{z}$ direction. Defining the plane of the substrate as
the $x-z$ plane, we let the $z$ axis lie exactly between the wires,
and let $y$ axis point to the region above the substrate. \cite{hinds}

We then proceed with the addition of two bias fields, 
one in the $\hat{z}$ direction, $B_{bz}$, and one in the  $\hat{x}$
direction, $B_{bx}$. The first of the two is put in place in order to avoid
regions of exactly zero field, where Majorana transitions would cause
arbitrary spin flips, and therefore loss of atoms from the guide.
The second of the two fields, when added vectorially to the field
generated by the wires, generates regions of minimum potential in the  $x-z$
plane. In particular, for $B_{bx}^0=\mu_0 I / 2\pi d$ ,where $\mu_0$ is
the permeability of free space, and $d$ is the separation between the
wires, there exists a single potential minimum located on the y axis
a distance $y_0 = d$ above the wires.

Furthermore, for $B_{bx} < B_{bx}^0$ two minima are generated on the y
axis, one above and one below  $y_0$, and for $B_{bx} > B_{bx}^0$, two
minima are again generated above the substrate, but this time they are
displaced symmetrically to the left and the right of the $y-z$ plane.
It is this latter regime that we use to generate a beam splitter,
letting the wires be fixed, and changing the transverse bias field
$B_{bx}$ as a function of $z$ from $B_{max}$ to $B_{min}$ and back, such
that $B_{max} > B_{min} > B_{bx}^0$.

\begin{figure}[ht]
\includegraphics[width=9.0cm,height=7.0cm]{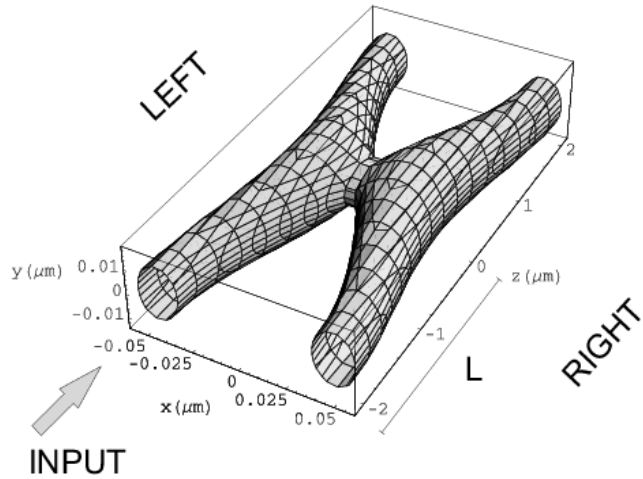}
 \caption{Constant Energy surface representing the potential in
  Eq.(\ref{btopot}). The parameters chosen for this plot are $d=0.1\mu
  m$, $B_{min}=21.3 G$, $B_{max}= 22.5 G$, $B_{bz}=1.0 G$, $L=20\mu m$. The
  surface contour is drawn at the energy of the 
  lowest mode of the input arm of the beamsplitter, and we define ``left''
  and ``right'' arms for labeling convenience.}
\label{3dpot} 
\end{figure}

The magnetic field produced by such configuration is therefore
\cite{zozulya2,hinds} 
 
\begin{eqnarray}
& B_x &= \frac{\mu_0 I}{2 \pi}\left[ \frac{-y}{(x-d)^2 + y^2}
 +\frac{-y}{(x+d)^2 + y^2}\right] + B_{bx}(z) \nonumber \\
& B_y &= \frac{\mu_0 I}{2 \pi}\left[ \frac{x-d}{(x-d)^2 + y^2}
 +\frac{x+d}{(x+d)^2 + y^2}\right] \nonumber \\
& B_z &= B_{bz}  \nonumber \\
\label{bfields}
\end{eqnarray}
and the consequent potential experienced by the atoms is
\begin{equation}
V=g_F \mu_B m_F |\mathbf{B}|,
\label{btopot}
\end{equation}
where $\mu_B$ is the Bohr magneton, $g_F$ is the Land\'e factor,
$m_F$ is the total angular momentum projection quantum number, and the
atoms' spin is aligned with the field at every point in space. An
example of a guiding potential is illustrated in Fig \ref{3dpot}. 

The adjustable experimental parameters are therefore the current in
the wires $I$, the values of the bias fields $B_{max}$ , $B_{min}$ and $
B_{bz}$, and the distance $d$ between the wires. Throughout this work
we choose, for concreteness,  $d=0.1\mu m$, $B_{min}=21.3 G$,
$B_{max}= 22.5 G$, 
$B_{bz}=1.0 G$, $L=20\mu m$, and we let $B_{bx}(z)$ be a fourth order
polynomial in z, such that it has zero derivative at the center (
$z=0$) and
edges ($z=\pm L$) of the beam splitter. Also we will only consider cases in which
$B_{bx}(z)$ reaches its minimum value at $z=0$ only, avoiding the
characterization of the trivial evolution of the wave function at a
constant field. In particular we will consider the following form for
the variation of the transverse bias field:

\begin{equation}
B_{bx}(z)=B_{mim} +(B_{min}-B_{max})\left[ 2 (x/L)^2-(x/L)^4\right]
\label{bofz}
\end{equation}
Varying $L$ will therefore adjust the adiabaticity of the beamsplitter,
whose effects we will study in section (III B)
 
\subsection{Waveguides as a Scattering Problem}

Because we are going to treat the beamsplitter as a scattering problem, we
will begin by offering a quick review of scattering theory; in
particular we will reproduce the basic formulation of the adiabatic
treatment of the scattering problem. 

Scattering theory is fundamentally based on the superposition
principle, which constrains us to the solution
of the linear Schr\"{o}dinger equation. This limit is
nonetheless justifiable in light of the known problems caused by the
interaction between atoms, such as the wave function recombination
instabilities described in Ref.\cite{zozulya3}.

The separation between the guides at the input and output ports of 
the beam-splitter is sufficiently great that no tunneling is
possible between the guides within the time frame of the experiment.
The problem is thus divided into two separate regions. We will refer to the
region $|z|<L$ as the scattering region. This is the inner
region containing the active part of the beam-splitter, where all the
coupling between the modes takes place. In the outer region, defined
by $|z|>L$,the potential has translational symmetry in $z$. 
Solutions to the Schr\"{o}dinger equation in the outer
region are therefore trivially found to be products of transverse modes and
longitudinal plane-wave solutions. The problem is thus reduced to finding
solutions inside the scattering region, and match them to the
solutions outside to find solutions to the global problem. Once these
solutions are found it is then possible to generate the S-matrix
for the system. 

Moreover, since we are matching at the boundary of the scattering
region the only information we need is the value of the wave
function and its derivative at the boundary, and nowhere
else. In particular we need to compute the quantity
\begin{equation}
b=-\left. \frac{1}{\Psi(\mathbf{r})} 
\frac{\partial\Psi(\mathbf{r})}{\partial n}\right|_{\Sigma},
\label{logder}
\end{equation}
defined as the logarithmic derivative,
where $\Sigma$ is the boundary of the scattering region, $n$ 
is the outward normal to the surface $\Sigma$, and $\Psi$ is the
wave function in the inner region. Because the wave function must vanish
in the limit of large $|x|$ or $|y|$, the surface $\Sigma$ consists,
for us, of the two planes $z=\pm L$. To find solutions inside the box we
used the R-matrix method, 
 formulated in the adiabatic representation. A derivation of this
 method follows. 

\subsection{The Adiabatic representation}

We start by writing the Shr\"{o}dinger equation
\begin{equation}
-\frac{\hbar^{2}}{2 m}\frac{\partial^2}{\partial z^2}\Psi(x,y,z) +
 \left[ -\frac{\hbar^{2}}{2 m} \nabla_{x,y}^{2} +V(x,y,z) \right]
 \Psi(x,y,z) = E \Psi(x,y,z), 
\label{shrod}
\end{equation} 
with $V(x,y,z)$ as defined in Eq.(\ref{btopot}).
If atoms in the guide were moving infinitely slowly ,
i.e. adiabatically, then the wave function would be well represented by
the basis set $\phi_i(x,y;z)$ with eigenvalues 
$U_i(z)$, defined as solutions to the equation
\begin{equation}
\left[ -\frac{\hbar^{2}}{2 m} \nabla_{x,y}^{2} +V(x,y;z) \right]
\phi_i(x,y;z) =  U(z)\phi_i(x,y;z). 
\label{spect}
\end{equation} 
As in the Born-Oppenheimer approximation, the quantities $U_i(z)$
serve as effective potentials for the subsequent motion in $z$.
To recover the effect of finite velocity in $z$, it would be
appropriate to expand the wave function in 
terms of the adiabatic basis in the following way:
\begin{equation}
\Psi(x,y,z)=\sum_{j}F_j(z)\phi_j(x,y;z),
\label{psiexp}
\end{equation}
where the $z$ dependence of the coefficient $F_i(z)$ is necessary in
order to restore the motion in the $z$ coordinate.
We should note that the above defined basis functions depend only
parametrically on $z$, and they are normalized  in the following way: 
\begin{equation}
\int \phi_i(x,y;z)\phi_j(x,y;z) dx dy = \delta_{i,j}.
\label{basisnorm}
\end{equation}
This normalization implies that all transverse functions must vanish
as $x,y \rightarrow \infty$, and therefore defines the effective boundary 
of the scattering region as $z= \pm L$.

Having defined the basis set we proceed to insert Eqn. (\ref{psiexp}) into
Eqn. (\ref{shrod}), and subsequently project the resulting equation onto
$\phi_i(x,y;z)$, to obtain the set of coupled equations
\begin{equation}
-\frac{\hbar^{2}}{2 m}\left[\frac{\partial^2}{\partial z^2} F_i(z) +
2\sum_{j}P_{ij}\frac{\partial}{\partial z} F_j(z)+
\sum_{j}Q_{ij}F_j(z)\right] +U_i(z)F_i(z)=E F_i(z) 
\label{adiashrod}
\end{equation} 
where we have defined, as conventional, 
\begin{eqnarray}
&P_{ij}=&\left<\phi_i(x,y;z)\left|\frac{\partial}{\partial
  z}\phi_j(x,y;z)\right>\right.
\label{pdef}\\
&Q_{ij}=&\left<\phi_i(x,y;z)\left|\frac{\partial^2}{\partial
  z^2}\phi_j(x,y;z)\right>.\right.
\label{qdef}
\end{eqnarray}
$P$ and $Q$ are operators of momentum-like and kinetic energy-like
quantities, and thus reflect the influence of finite propagation
velocity in $z$.
Notice that $P$ and $Q$ vanish by construction in the outer region.
 
We have thus cast the original 3-dimensional problem into a collection
of 2-dimensional problems to find $\phi_i(x,y;z)$ and 
$U_i(z)$, and a 1-dimensional coupled channel problem to find
$F_i(z)$. The advantages of this shift in paradigm are twofold:
on the one hand, a very complicated and computationally lengthy problem is
turned into a simpler and computationally manageable problem;
On the other hand, the adiabatic approach lends itself very naturally to
approximations and qualitative understanding of the underlying
physics.  

\subsection{The R-Matrix Method}

As mentioned earlier, solving the scattering problem implies finding
the logarithmic derivative $b$ as defined in Eq.(\ref{logder}). In atomic
structure physics, there is a well known variational principle for the
logarithmic 
derivative, which follows rather simply from the Schr\"{o}dinger
equation \cite{greene}: 
\begin{equation} 
b=\frac{\int_{\Omega} \left[\vec{\nabla}\Psi^*\cdot\vec{\nabla}\Psi
    +\frac{2m}{\hbar^{2}}\Psi^* (V-E)
    \Psi\right]}{\int_\Sigma \Psi^*\Psi}, 
\label{varb}
\end{equation}
where $\int_{\Omega}$ denotes an integral over the volume of the scattering
region, while $V$ is the scattering
potential, and $\int_\Sigma$ is a surface integral over the surface
bounding the scattering region.

The typical approach to the problem, at this point is to
expand the wave function in a complete set of basis functions
${Y_k}$, to get 
$\Psi(x,y,z)=\sum_k C_k Y_k(x,y,z)$, and
take matrix elements of the operators in Eqn. (\ref{varb}) with respect to
such basis, to obtain the following generalized eigenvalue
problem\cite{greene}:
\begin{equation}
\hat{\Gamma} \vec{C} = b \hat{\Lambda} \vec{C},
\label{eig1}
\end{equation}
where
\begin{eqnarray}
\Gamma_{ij}&=&\int_{\Omega} \left[\vec{\nabla}Y_i^*\cdot\vec{\nabla}Y_j
    +\frac{2m}{\hbar^{2}}Y^*_i (V-E) 
    Y_j\right]
\label{gamma1}\\
\Lambda_{ij}&=&\int_{\Sigma}Y_i^*Y_j.
\label{lambda1}
\end{eqnarray}
This is the form that the eigenchannel R-matrix takes in a diabatic
representation.

The solutions of the eigenvalue problem consists of a set of
eigenvectors $C^{\sigma}$, and a set of
eigenvalues $b^{\sigma}$, representing the logarithmic derivatives of the
functions $\Psi^{\sigma}=\sum_k C_k^{\sigma} Y_k$. The newly introduced
index $\sigma$ refers to the different possible internal states of the
system, called R-matrix eigenchannels. 
The concept of eigenchannel in scattering theory can be understood by
analogy with the concept of eigenstate in bound state problems. In
fact, just like the energy variational principle leads to an
eigenvalue problem for bound state eigenfunction and corresponding
energies, the variational principle in Eq.(\ref{varb}) leads to a set
of eigenchannels, with corresponding eigen-logarithmic derivatives.  

As we mentioned earlier, we plan to work using the adiabatic basis
defined in Eq.(\ref{psiexp}),  
so we expand Eq.(\ref{varb}) in terms of this set, and obtain the
following variational principle:
\begin{equation}
b=\frac{\sum_{ij}\int dz \left[\left\{\frac{\partial}{\partial z}
    F_i^*\frac{\partial}{\partial z} F_j +\frac{2m}{\hbar^{2}}
    F_i^*(U_i(z)-E)F_j \right\}\delta_{ij}+
2F_i^*P_{ij}\frac{\partial}{\partial z} F_j+
F_i^* Q_{ij}F_j
    \right]}{|F_i(\Sigma)|^2}. 
\label{varb2}
\end{equation}

Since the adiabatic basis only defines motion in the transverse
coordinates, it remains to expand the longitudinal functions $F_i(z)$
with an arbitrary set of $z$ dependent functions, in our particular
case we chose basis-splines,  in the form 
$F_i(z)=\sum_k c_{ik} y_k^i(z)$. We can now write the adiabatic
equivalent to    
Eq.(\ref{eig1}). 

In order to simplify the notation, we combine the
indices $i,k$ into 
the index $\alpha$, so that $c_{\alpha}$ becomes a vector, and we write
\begin{equation}
\hat{\Gamma}^a \vec{c} = b \hat{\Lambda}^a \vec{c},
\label{eig2}
\end{equation}
where
\begin{eqnarray}
\Gamma^a_{\alpha\beta}&=&\int dz \left[\left\{\frac{\partial}{\partial z}
    y_{\alpha}^*\frac{\partial}{\partial z} y_{\beta} +\frac{2m}{\hbar^{2}}
    y_{\alpha}^*(U_i(z)-E) y_{\beta}\right\}\delta_{ij}+
y_{\alpha}^*P_{ij}\frac{\partial}{\partial z}y_{\beta} +
\frac{\partial}{\partial z}y_{\alpha}^*P_{ji}y_{\beta} +
y_{\alpha}^*Q_{ij}y_{\beta}
    \right]
\label{gamma2} \nonumber \\
\Lambda^a_{\alpha\beta}&=& y_{\alpha}^*(\Sigma)y_{\beta}(\Sigma)\delta_{ij}.
\label{lambda2}
\end{eqnarray}

In the above equations we have written the P-matrix portion of
Eq.(\ref{varb2}) in an Hermitian form by integrating by parts, and setting
the resulting surface integral to zero, using the fact that by
definition all couplings $P,Q$ must vanish outside the scattering region.

\subsection{The Outer Region: Matching and Physical Consideration}

Having solved Eq.(\ref{eig2}), one obtains a set of eigenvalues
$b^{\sigma}$, and a 
set of eigenvectors $\vec{c^\sigma}$. 
It therefore follows that on the boundaries $\Sigma$  of the
scattering region we can connect the inner and outer solutions by: 

\begin{eqnarray}
&&\Psi^{\sigma}(x,y,\Sigma)=\sum_{j}F_{j}^
{\sigma}(\Sigma)\phi_{j}(x,y,\Sigma)=
\sum_{\alpha}c_{\alpha}^{\sigma} y_{\alpha}(\Sigma) \phi_{j}(x,y,\Sigma)
\nonumber\\
&&=
\sum_{\alpha}\phi_{j}(x,y,\Sigma)*\left(A_{j}^{\sigma}\frac{e^{-ik_{j}\Sigma}}
    {\sqrt{2|k_i|}}
 + B_{j}^{\sigma} \frac{e^{ik_{j}\Sigma}}{\sqrt{2|k_i|}}
\right)
\nonumber\\
\label{bbb}
\end{eqnarray}
where $k_i=\sqrt{2m(E-U_i(\Sigma))}$ is real for $E>U_i(\Sigma)$,
and imaginary 
for $E<U_i(\Sigma)$, and $\Sigma = \pm L$. At a particular incoming
energy, we define a channel  
with real $k_i$ to be ``open'' (meaning energetically available) an channel 
with imaginary $k_i$ to be ``closed''. If a channel $i$ is closed we set
$A_{j}^{\sigma}=0$, to avoid unphysical divergences.   
A similar argument is valid for the derivative of the wave function:
\begin{eqnarray}
&&\frac{\partial}{\partial \Sigma
  }\Psi^{\sigma}(x,y,\Sigma)=-b^{\sigma}\Psi^{\sigma}(x,y,\Sigma)
\nonumber\\
\label{aaa}
\end{eqnarray}
Eqs(\ref{bbb},{\ref{aaa}), together with the orthonormality of the set
${\phi_{i}}$, and the assumption of unit incoming flux,
imply that $F_{j}^{\sigma}(\Sigma)$  and its derivatives can be written as a linear
combination of  the form:

\begin{eqnarray}
\left\{
\begin{array}{rl}
F_{i}^{\sigma}(\Sigma) &=
\left[\delta_{ij}\frac{e^{-ik_{i}\Sigma}}{\sqrt{2|k_i|}}
 - S_{ij} \frac{e^{ik_{i}\Sigma}}{\sqrt{2|k_i|}}\right]
   N_{j}^{\sigma} \\ \\ 
b^{\sigma}F_{i}^{\sigma}(\Sigma) &= -\left[ \delta_{ij}
\frac{ik e^{-ik_{i}\Sigma}}{\sqrt{2|k_i|}}
 - S_{ij} \frac{-ike^{ik_{i}\Sigma}}{\sqrt{2|k_i|}}\right]
   N_{j}^{\sigma} \\
\end{array} 
\right.
\end{eqnarray}

The quantity
$N_{i}^{\sigma}$ is a factor which serves to connect
the normalizations of the two equations. On the other hand
$S_{ij}$ is the scattering matrix of the system (often referred to as
S-matrix), and it represents the probability amplitude to enter the
beam splitter in 
channel $j$, and exit it in channel $i$, or vice versa, since $S$
is Hermitian due to time reversal symmetry. 

Moreover, since the
equation is true on the whole of the boundary, the channel index
describes the probability amplitude for the atom to be found at either
end of the beam splitter (in fact at any particular arm of the beam
splitter), in some particular mode. This allows us to
calculate mode mixing, as well as reflection and transmission
amplitudes.The above system of equations can be solved for the
unknowns  $S_{ij}$ and $N^{\sigma}_j$.

\subsection{Solving the Equations: Considerations on Numerical and
  Mathematical Details}

The numerical problem consists of two main parts. The first is to find the
transverse eigenmodes $\phi_i(x,y;z)$. This is accomplished by solving
Eq.(\ref{spect}) at various values of $z$  in such a way that the adiabatic
curves $U_{i}(z)$ may be interpolated easily. We accomplish this task
by generating a Hamiltonian matrix, again using b-splines as a basis
set, and diagonalizing it at various values of $z$.

Furthermore one needs to evaluate the $P$ and $Q$ matrices in Eqs.(
\ref{pdef},\ref{qdef}). To do this one may exploit the Hellmann-Feynman
theorem to obtain the following expressions: \cite{child}
\begin{equation}
 P_{ij}(z)=\left\{ \begin{array}{c c}
\frac{-M_{ij}}{U_{i}(z)-U_{j}(z)} & i \neq j \\ 0 & i=j 
\end{array}\right.,
\label{hfp}
\end{equation}
and
\begin{equation}
 Q_{ii}=- \sum_{k \neq i} \frac{M_{ik}M_{ki}}{U_i(z)-U_k(z)},
\label{hfq}
\end{equation}
 where
\begin{equation}
 M_{ij}= \int \phi_{i}^*(x,y,z) \left\{ \frac{\partial}{\partial z}
  V(x,y,z)\right\} \phi_{j}(x,y,z) \ dx dy ,
\end{equation} 
 We adopt a common approximation whereby $Q_{ij}=0$ for $j \neq
i$.
The second part consists of a scattering problem on the adiabatic
curves, by choosing a basis set $y_i(z)$. For our calculations we use
b-splines.\cite{deboor,hart} 

The guiding potential $V(x,y,z)$ in Fig. \ref{3dpot}, exhibits a reflection
symmetry about the $x-y$ plane. Such a symmetry implies that there is
no coupling between even and odd transverse modes of the beam
splitter. This in turn implies that by describing the problem in a basis of
even-odd modes it is possible to solve two smaller problems,
significantly reducing the computational effort. At
the end of the calculation it is then possible to perform a change of
basis to a ``left-right'' set describing the ``left'' and ``right'' arms of the
beamsplitter, where ``right''=``even-odd'', and ``left''=``even+odd.''

\section{Results}

Having described the general formalism, we proceed to report some
quantitative results. In particular, we use the parameters described in
the caption of Fig. \ref{3dpot}, and study the behavior of the system
as we vary the length $L$ over which the beam is split. We focus
especially on the non adiabatic 
characteristics of the beamsplitter, namely reflection and higher mode
excitation. 

The parameters that generate the guiding potential in our model are
consistent with those in recent chip-based experiments
\cite{leanhardt,schwindt}. The major difference is that our model guides lie
close to the substrate, thus tightly confining the atoms in the
transverse direction. At reasonable atom velocities of several cm/sec, 
only two modes are then energetically open, simplifying the calculations and
interpretation in this pilot study. More realistic beamsplitters can
be handled by including the appropriate number of modes in the calculation.

\subsection{The Adiabatic Curves}

The simplest level of approximation for the problem is to consider
only the first even and odd mode of the structure, and,
analogously to the Born-Oppenheimer approximation, ignore all higher
modes and couplings. 
Within the framework of such an approximation we see that the
Born-Oppenheimer potential depends only on the transverse frequency of the
guide, which is highest at the entrance and exit of the beamsplitter and
lowest in the center, giving rise to curves resembling smoothed square
wells. 
As it turns out the predictions of
this simple model prove to be grossly inadequate when compared to full
coupled channel calculation.  The reason for this is that the 
Born-Oppenheimer channels are strongly coupled by nonadiabatic effects.

To suggest how big a correction nonadiabatic effects are, we compare the
lowest-lying Born-Oppenheimer potential $U_0(R)$ (dashed line in Fig. 2) 
to the so-called
``adiabatic'' potential, defined by $U^{eff}_0(z)=U_0(z)+Q_{00}(z)$
(solid line).  The $Q_{00}(z)$ term represents an effect of the 
transverse momentum on the longitudinal motion. As the guiding potential 
varies as a function of $z$, the paths of the atoms
follow the centers of the guides. 
This causes the atoms to acquire transverse momentum, which
removes kinetic energy from the longitudinal motion.  Thus $Q_{00}$
is a positive correction.  

In chemical physics applications, the
adiabatic curve is sometimes, but not always, a better single-channel
representation of the problem \cite{Sucre}.  In our case, it usefully 
incorporates a primary effect arising from nonadiabaticity.
Namely, $U_{0}^{eff}$ possesses a barrier at the input of the splitter. 
This barrier reflects  the fact that kinetic energy spent in transverse
motion halts motion in the longitudinal direction.  Effects of this barrier
are evident in the fully-converged scattering calculations, below. 

A more complete set of effective adiabatic curves for the first few even
and odd modes is shown in Fig.\ref{adia}.  For kinetic energies greater
than $\sim 20 \mu$K, excited state potentials are energetically allowed 
in the scattering region.  The corresponding mode mixing
can be thought of as the ``sloshing'' of the
condensate as it is pulled side to side in the potential. 
Even if these excited channels are not energetically allowed, they
may (and do) still perturb propagation in the lowest mode.
Since the length $L$ of the beamsplitter is, in our
model, thousands of times larger than the longitudinal de Broglie
wavelength of the atoms, even a small coupling between channels can
cause a drastic change in phase shift. This implies that we need a fully
coupled channel calculation to solve the problem quantitatively.

\begin{figure}[ht]
\includegraphics[width=9.0cm,height=7.0cm]{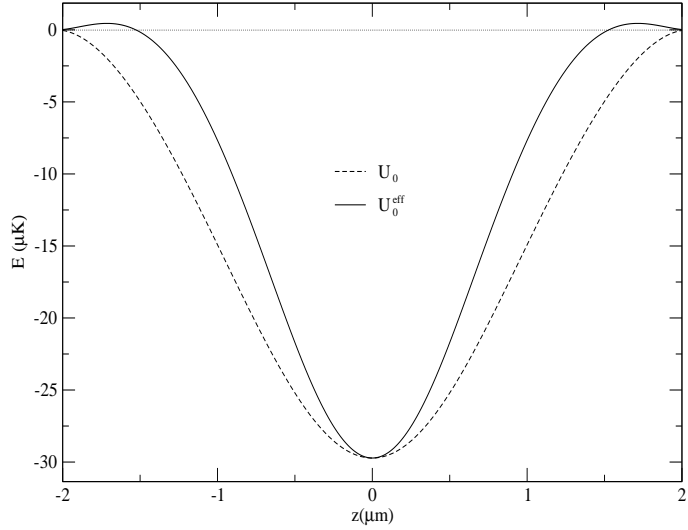} 
\caption{Comparison of lowest-lying Born-Oppenheimer (U$_0$) and adiabatic
  (U$^{eff}_0$) curves, for the beamsplitter in Fig. \ref{3dpot}}
\label{born} 
\end{figure} 

\begin{figure}[ht]
\includegraphics[width=9.0cm,height=7.0cm]{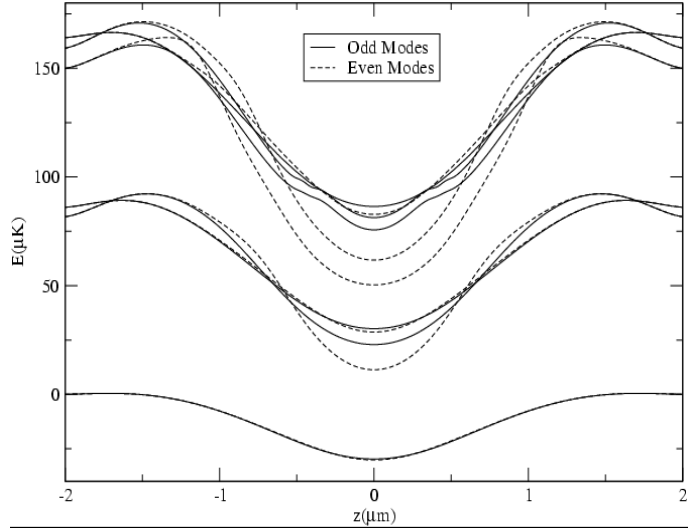} 
\caption{Effective adiabatic potentials for the configuration in
  Fig. \ref{3dpot}. Each curve corresponds to a different transverse
  mode of the beamsplitter. Because of the intrinsic symmetry of the
  potential, even and odd modes can be treated
  separately. The first six even and odd modes of the structure are
  depicted here.}
\label{adia} 
\end{figure}

\begin{figure}[ht]
\includegraphics[width=9.0cm,height=7.0cm]{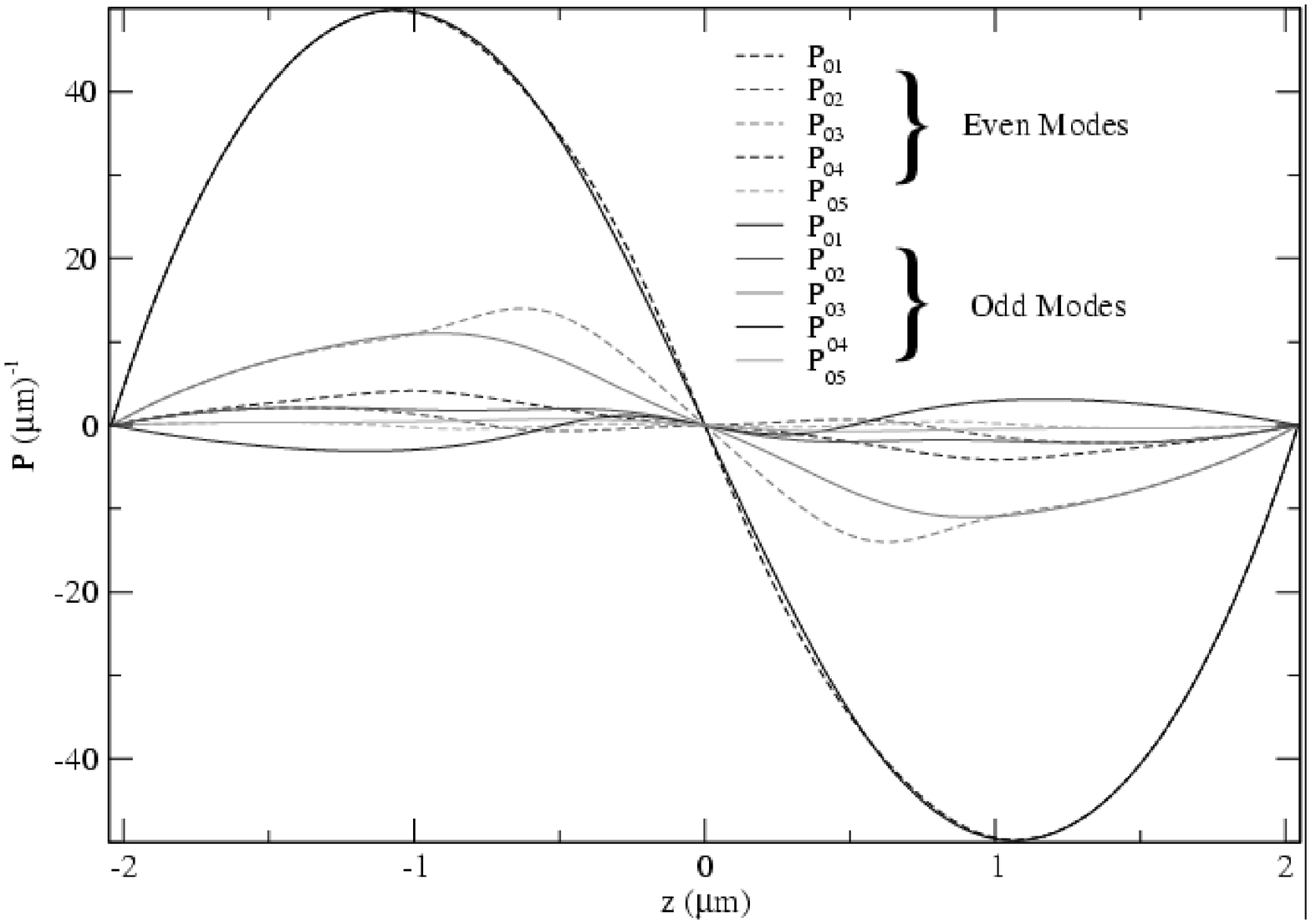} 
\caption{P-matrix elements coupling the first transverse even and odd
  modes in Fig. \ref{adia} to selected higher ones.}
\label{pmat} 
\end{figure} 

Channel coupling is achieved through off-diagonal elements of the
$P$ matrix, several of which are shown in Fig.\ref{pmat}.
As expected, the couplings $P_{0i}$ between the lowest channel 0
and higher channels $i$ diminishes as $i$ gets larger.
Also, as implied by Eq.(\ref{hfp})
the coupling is strongest where the potential is steepest in the
longitudinal direction (i.e. $z= \pm 1$ in the figure).

\subsection{General Features of Scattering}

Having defined the terms of the problem, and calculated the adiabatic
curves and couplings, we solve the scattering equations and extract
the S-matrices of the system. All figures shown to this point refer to
a beamsplitter with $L=20 \mu m$, which is one in which most of the
typical features are present. Fig. \ref{l=20} shows the absolute
values of selected S-matrix elements for this configuration,
which represent the probabilities for various outcomes.
In particular we show the probabilities to exit in the
various arm of the beam splitter, assuming unit input flux from the
left arm of the splitter, as defined in Fig. \ref{3dpot}. 
At the incident energies shown in Fig. \ref{l=20}, only the lowest mode
in each arm is energetically accessible.
This typical case is illustrative of the basic elements of the
beamsplitter. 

\begin{figure}[ht]
\includegraphics[width=9.0cm,height=7.0cm]{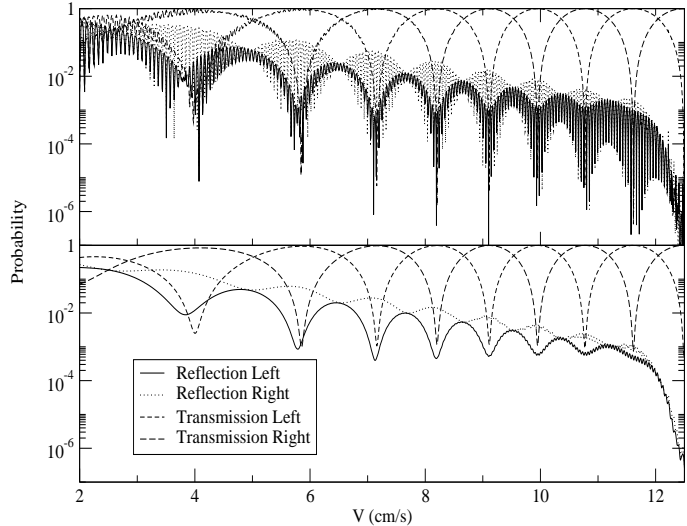} 
\caption{Probability for exit in the various arms of the beamsplitter
  depicted in Fig. \ref{3dpot}, versus incident atom velocity.
  This figure assumes that the atoms have entered in the lowest mode
  of the left arm.  The velocities shown correspond to atom energies
  below the second threshold, thus suppressing higher mode excitations. The
  bottom panel is a $26 \mu K$ Gaussian convolution of the 
  one above. }
\label{l=20} 
\end{figure} 

In this beamsplitter the largest probabilities (dashed and short-dashed
lines in Fig. \ref{l=20}) correspond to transmission, with the
probability alternating between left and right arms.  Thus approximately
50-50 beamsplitting is possible at atom energies where these two
curves cross.  Moreover, the sum of the left and right transmission
probabilities is almost, but not quite, equal to unity.  This can
be seen in the slowly decreasing reflection probabilities (solid and
dotted lines) in the figure.  The general features of beamsplitting
are preserved under a convolution in energy, as exhibited in Fig.
\ref{l=20} b).  Here and in what follows, convolution is used to 
simplify the appearance of the calculations.

The reflection probabilities also exhibit a similar left-right oscillation
as a function of energy. In addition, they exhibit a much faster
oscillation.  This faster oscillation is familiar from one-dimensional
scattering from a potential, with one oscillation being added each time
the energy increases enough to introduce a new de Broglie wavelength
into the scattering region \cite{Gas}.  Here the oscillations are
numerous, since the guiding potential is thousands of de Broglie wavelengths
long.  (These oscillations are of course
also present in the transmission probabilities, but are too small to be
seen on the scale of the figure.)

For smaller values of $L$ , the beamsplitter is badly non-adiabatic,
and even qualitative features of beamsplitting fade. Fig. \ref{l=10} shows
such a non-adiabatic case, with $L = 1 \mu m$. The effect of the input barrier 
described in Fig. \ref{born}, is now much larger, suppressing all
transmission up to input velocities of about 5cm/s. As the kinetic
energy reaches the energy of the barrier, the probability exhibits
resonant behavior by the presence of spikes in the 
S-matrices. Though mostly washed out by convolution, these features
would in principle cause transparency of the barrier at extremely well 
defined velocities, where the kinetic energy equals the energy of a 
metastable boundstate.  At higher atom velocities, above the input barrier,
reflection remains extremely likely, and even the basic action of the
beamsplitter is destroyed.

\begin{figure}[ht]
\includegraphics[width=9.0cm,height=7.0cm]{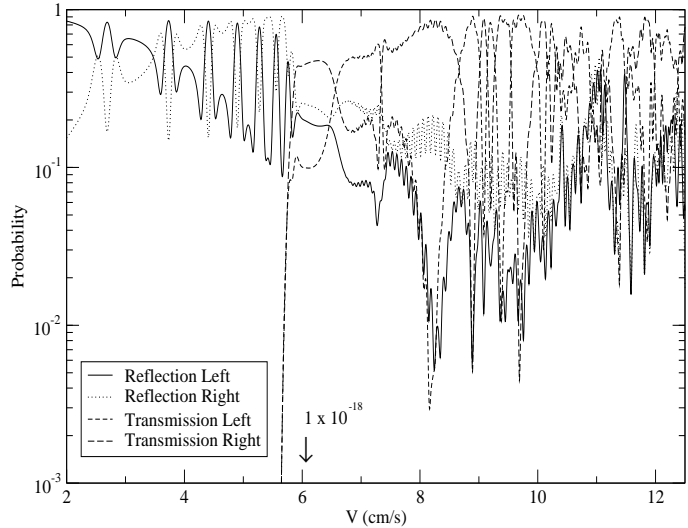} 
\caption{Scattering probabilities as in Fig. \ref{l=20}, but for the extremely
  non adiabatic case $L=1\mu m$. Only the convolved plot is shown,
  with a width of 26 $\mu$K.} 
\label{l=10} 
\end{figure}

\subsection{Towards Adiabaticity}

Fig. \ref{refl} shows reflection probabilities versus atom velocity,
for various values of the beam-splitter lengths $L$.
These results are convolved over an energy
width of $16 \mu K$, to emphasize the overall probability
rather than the oscillatory structure. For $L>2 \mu$m, reflection decreases 
nearly linearly on this semi-log plot, suggesting an exponential
decrease of reflection probability with velocity.
Reflection also decreases with increasing $L$, as expected for
an increasingly adiabatic beamsplitter.
 The features noticeable around
12.5cm/s  and 13.5cm/s represent cusps at the thresholds for the second and
third mode to become energetically available, smoothed out by convolution.

\begin{figure}[ht] 
\includegraphics[width=9.0cm,height=7.0cm]{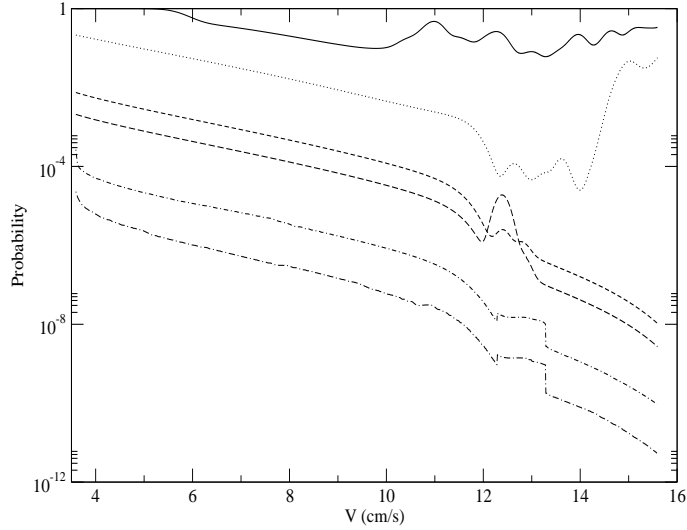} 
\caption{Total reflection probabilities for multi-mode beamsplitter
  of different lengths. From top to bottom $L=1,2,5,7,15,30 \mu
  m$. The cusps around 12.5 and 13 $cm/s$ are the effects of the
  second and third thresholds becoming energetically available. The
  plot represents a $16 \mu K$ width Gaussian convolution.} 
\label{refl} 
\end{figure} 

Finally, in Fig. \ref{trans} we plot the total transmission to modes higher
than the first, for input velocities higher than the second mode
threshold. As might be expected, the probability to generate higher
modes grows as a function of atom velocity.  Countering this trend,
the probability again diminishes as the length $L$ becomes longer.
\begin{figure}[ht]
\includegraphics[width=9.0cm,height=7.0cm]{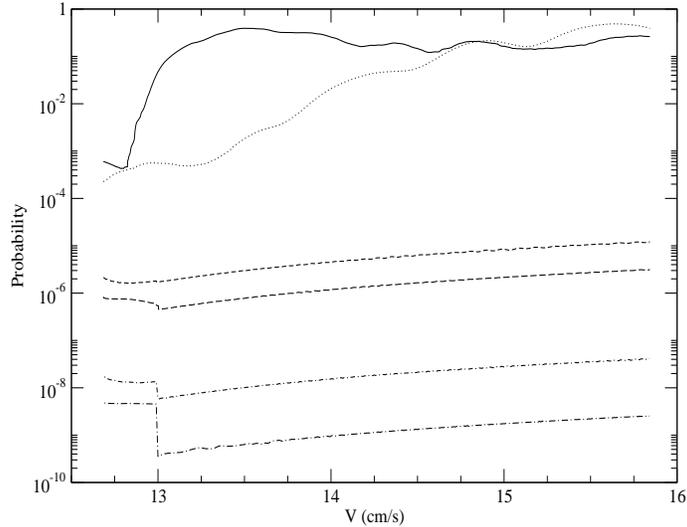} 
\caption{Total transmission to higher modes for different lengths
  beamsplitter. From top to bottom $L=1,2,5,7,15,30 \mu
  m$.  The plot represents a $3.25 \mu K$ width Gaussian convolution.} 
\label{trans} 
\end{figure} 

\section{Conclusions}

We have developed a novel approach to the analysis of
non-interacting atomic beams traveling through waveguides, 
based on the adiabatic representation of scattering theory.   
This method, originally developed for the study of molecular collision
theory, is known to be very flexible, and  could be
applied to many other guiding geometries.
We applied this approach to the study of a two wire atomic
beam-splitter, both to illustrate the method and
to explore a particular guiding geometry. We have found that the
nonadiabatic couplings play a significant role.  Because
we have deliberately focused on a tightly-confining geometry,
it is likely that nonadiabaic effects are even more significant in
realistic beamsplitters.  This will be a topic of future study.

\acknowledgements
This work was supported by a grant from  ONR-MURI.

\end{document}